\documentclass[conference]{IEEEtran}
\usepackage{color}
\usepackage{multicol}
\usepackage{graphicx}
\usepackage{multirow}
\usepackage{booktabs}
\usepackage[table]{xcolor}
\usepackage{algorithm}
\usepackage{algorithmic}
\usepackage{amsmath,amssymb,amsthm}
\usepackage{epic,multibox,fancybox}
\usepackage{float}
\usepackage{blindtext}

\usepackage{subfigure}
\usepackage{graphicx}
\usepackage{caption}
\ifCLASSINFOpdf

\else

\fi
\begin{document}

\title{SIMPLE: Stable Increased-throughput\\ Multi-hop Protocol for Link Efficiency in\\ Wireless Body Area Networks}

\author{Q. Nadeem$^{1}$, N. Javaid$^{1,2}$, S. N. Mohammad$^{1}$, M. Y. Khan$^{1}$, S. Sarfraz$^{1}$, M. Gull$^{3}$\\\
$^{1}$Dept of Electrical Engineering, COMSATS Institute of IT, Islamabad, Pakistan.\\
$^{2}$CAST, COMSATS Institute of IT, Islamabad, Pakistan.\\
$^{3}$NWFP, UET, Peshawar, Pakistan.}
\maketitle
\begin{abstract}
In this work, we propose a reliable, power efficient and high throughput routing protocol for Wireless Body Area Networks (WBANs). We use multi-hop topology to achieve minimum energy consumption and longer network lifetime. We propose a cost function to select parent node or forwarder. Proposed cost function selects a parent node which has high residual energy and minimum distance to sink. Residual energy parameter balances the energy consumption among the sensor nodes while distance parameter ensures successful packet delivery to sink. Simulation results show that our proposed protocol maximize the network stability period and nodes stay alive for longer period. Longer stability period contributes high packet delivery to sink which is major interest for continuous patient monitoring.

\end{abstract}

\IEEEpeerreviewmaketitle
\section{Introduction}
Wireless Sensor Networks (WSNs) are used to monitor certain parameters in many applications like environment monitoring, habitant monitoring, ~\cite{mainwaring2002wireless} battle field, agriculture field monitoring and smart homes. These wireless sensors are dispersed in sensing area to monitor field. WBAN is new emerging sub-field of WSN. A key application of WBAN is health monitoring. Wireless sensors are placed on the human body or implanted in the body to monitor vital signs like blood pressure, body temperature, heart rate, glucose level etc. Use of WBAN technology to monitor health parameters significantly reduces the expenditures of patient in hospital. With the help of WBAN technology, patients are monitored at home for longer period. Sensors continuously sense data and forward to medical server.\\
\indent In WBANs, sensor nodes are operated with limited energy source. It is required to use minimum power for transmitting data from sensor nodes to sink. One of the major obstacles in WBAN is to recharge the batteries. An efficient routing protocol is required to overcome this issue of recharging batteries. Many energy efficient routing protocols are proposed in WSN technology~\cite{javaid2013energy1},~\cite{manzoor2013q},~\cite{javaid2013wireles}. However, WSNs and WBANs have different architectures, applications and operate in different conditions. It is impossible to port WSN routing protocols to WBAN. Therefore, energy efficient routing protocol for WBAN is required to monitor patients for longer period.\\
\indent We propose a high throughput, reliable and stable routing protocol for WBAN. We deploy sensor nodes on the body at fixed places. We place sink at waist. Sensors for ECG and Glucose level are placed near the sink. Both these sensors have critical data of patient and required minimum attenuation, high reliability and long life therefore; these sensors always transmit their data directly to sink. Other sensors follow their parent node and transmit their data to sink through forwarder node. It saves energy of nodes and network works for longer period.\\
The rest of the paper is organized in following order. In section 2, we review related work, while Section 3 describes motivation for this work. Radio model is presented in section 4, while detail of the SIMPLE protocol is presented in section 5. Performance metrics and simulation results are presented in section 6 and 7 respectively. Finally, section 8 gives conclusion.

\section{Related Work}
In WBAN technology, large numbers of routing schemes are proposed. In this section, we present some proposed routing protocols.
In~\cite{javaid2012m} the author presented a thermal aware routing protocol. Each node selects a minimum hop rout to sink. When a parent node gets heated, the children nodes select another optimal route.\\
\indent Latre \emph{et al}. proposed A Secure Low-Delay Protocol for Multi-hop Wireless Body Area Networks (CICADA) routing protocol which consists of a spanning tree structure~\cite{latre2007low}. CICADA used Time Division Multiple Access (TDMA) protocol to schedule transmission of nodes. The nodes near the root act as forwarder nodes or parent nodes, these nodes collect data from their associated children nodes and relay to sink. Due to extra traffic load of children nodes on parent nodes causes parent nodes to deplete their energy fast.\\
\indent In~\cite{quwaider2009body} Quwaider \emph{et al}. presented a routing protocol which tolerates to changes in network. They used store and forward mechanism to increase the likelihood of a data packet to reach successfully to sink node. Each sensor node has the capability to store a data packet. In source to destination route, each node stores data packet and transmits to next node. Storing a data packet and then retransmitting causes more energy to consume and longer end to end delay.\\
\indent Ehyaie \emph{et al}.~\cite{ehyaie2009using} proposed a solution to minimize energy consumption. They deploy some non-sensing, dedicated nodes with additional energy source. This technique minimizes energy consumption of nodes and enhances the network lifetime, however, additional hardware required for relay nodes increase the cost of the network.\\
\indent A clustering based protocol A Self-Organization Protocol for Body Area Networks (ANYBODY) is proposed in~\cite{watteyne2007anybody}. The objective of this protocol is to restrict the sensor nodes to transmit data direct to sink. It improves the efficiency of network by changing the selection criteria of CHs.\\
\indent In~\cite{nabi2010robust} Nabi \emph{et al}. proposed a protocol similar to store and forward mechanism. They integrate this store and forward scheme with Transmit Power Adaption (TPA). To control transmission power consumption, all nodes know their neighbours. Nodes transmit data with minimum power and with a stable link quality.\\
\indent A similar method was proposed by Guo \emph{et al}. in~\cite{guo2010packet}. They also used Transmission Power Control (TPC) scheme as Nabi \emph{et al}. proposed. When link quality of a node decreased, an Automatic Repeat Request (ARR) is transmitted and retransmit the drop packet. Retransmission of lost packet increases the throughput of the network with the expense of energy consumption. \\
\indent Tsouri \emph{et al}. in~\cite{tsouri2011investigation},~\cite{sapio2010low} used creeping waves to relay data packet. They proposed this protocol to minimize energy consumption of nodes while keeping the reliable on body link.\\
Authors in~\cite{javaid2013analyzing},~\cite{javaid2013analyzing2} analyze delay in WBANs and different medium access techniques for WBAN. In~\cite{quwaider2009probabilistic} author proposed a delay tolerant protocol. They compare their protocol with different existing protocols.

\section{Motivation}
Wireless Body Area Sensors are used to monitor human health with limited energy resources. Different energy efficient routing schemes are used to forward data from body sensors to medical server. It is important that sensed data of patient reliably received to medical specialist for further analysis. In~\cite{maskooki2011opportunistic} author presented a opportunistic protocol. Proposed scheme facilitate mobility at cost of low throughput and additional hardware cost of relay node. They deploy sink at wrist. Whenever sink node goes away from transmission range of nodes, it uses a relay node which collect data from sensor nodes. In opportunistic protocol, whenever patient moves his hands, the wireless link of sink with sensor nodes disconnects. Link failure consumes more power of sensor nodes and relay node also more packets will drop, which causes important and critical data to loss.\\
To minimize energy consumption and to increase the throughput, we propose a new scheme. Our contribution includes:\\
\begin{itemize}
\item Our proposed scheme achieves a longer stability period. Nodes stay alive for longer period and consume minimum energy.
\item Large stability period and minimum energy consumption of nodes, contribute to high throughput.
\end{itemize}

\section{Radio Model}
Many radio models are proposed in literature. We use first order radio model proposed in~\cite{heinzelman2000energy}. This radio model consider d, the separation between transmitter and receiver and d$^2$, the loss of energy due to transmission channel. First order radio model equations are given as.
\begin{equation*} %
E_{Tx}(k,d) = E_{Tx-elec}(k) + E_{Tx-amp}(k,d)
\end{equation*}
\begin{equation}\label{1}
E_{Tx}(k,d) = E_{Tx-elec}\times k + E_{amp}\times k\times d^2
\end{equation}

\begin{equation*}
E_{Rx}(k) = E_{Rx-elec}(k)
E_{Rx}(k) = E_{Rx-elec} \times k
\end{equation*}
\begin{equation}\label{2}
E_{Rx}(k) = E_{Rx-elec} \times k
\end{equation}

where E$_{Tx}$ is the energy consumed in transmission, E$_{Rx}$ is the energy consumed by receiver, E$_{Tx-elec}$ and E$_{Rx-elec}$ are the energies required to run the electronic circuit of transmitter and receiver, respectively. E$_{amp}$ is the energy required for amplifier circuit, while k is the packet size.\\
In WBAN, the communication medium is human body which contributes attenuation to radio signal. Therefore, we add path loss coefficient parameter ${n}$ in radio model. Equation 2 of transmitter can be rewritten as.
\begin{equation}\label{1}
E_{Tx}(k,d) = E_{elec}\times k + E_{amp}\times n \times k\times d^n
\end{equation}
The energy parameters given in equation 3 depend on the hardware. We consider two transceivers used frequently in WBAN technology. The Nordic nRF 2401A is a single chip, low power transceiver and other transceiver is Chipcon CC2420. The bandwidth of both transceivers is 2.4GHz. We use the energy parameter of The Nordic nRF 2401A transceiver because it consumes less power than Chipcon CC2420. The energy parameters for this transceiver are given in Table 1.\\
 \begin{table*}[t]
\centering
\begin{tabular}{|l|l|l|l|}
\hline
\textbf{Parameters} &\textbf{nRF 2401A} & \textbf{CC2420} & \textbf{Units}\\
\hline
DC Current(Tx)   & 10.5 & 17.4 & mA\\
\hline
DC Current(Rx)  & 18 & 19.7 & mA\\
\hline
Supply Voltage(min) & 1.9 & 2.1 & V \\
\hline
E$_{tx-elec}$ & 16.7 &96.9 & nJ/bit \\
\hline
E$_{rx-elec}$ & 36.1 &172.8 & nJ/bit\\
\hline
E$_{amp}$ & 1.97e-9& 2.71e-7 & j/b\\
\hline
\end{tabular}
\caption{Radio Parameters}
\label{tab:template}
\end{table*}

\section{SIMPLE: Protocol Detail}
In this section, we present a novel routing protocol for WBANs. The limited numbers of nodes in WBANs give opportunity to relax constraints in routing protocols. Keeping routing constrains in mind, we improve the network stability period and throughput of the network. Next subsections give detail of the system model and detail of SIMPLE protocol.
\subsection{System Model}
In this scheme, we deploy eight sensor nodes on human body. All sensor nodes have equal power and computation capabilities. Sink node is placed at waist. Node 1 is ECG sensor and node 2 is Glucose sensor node. These two nodes transmit data direct to sink.
Fig 1 shows the placement of nodes and sink on the human body.

\begin{figure*}[t]
\begin{center}
\includegraphics[height=14cm, width=12cm]{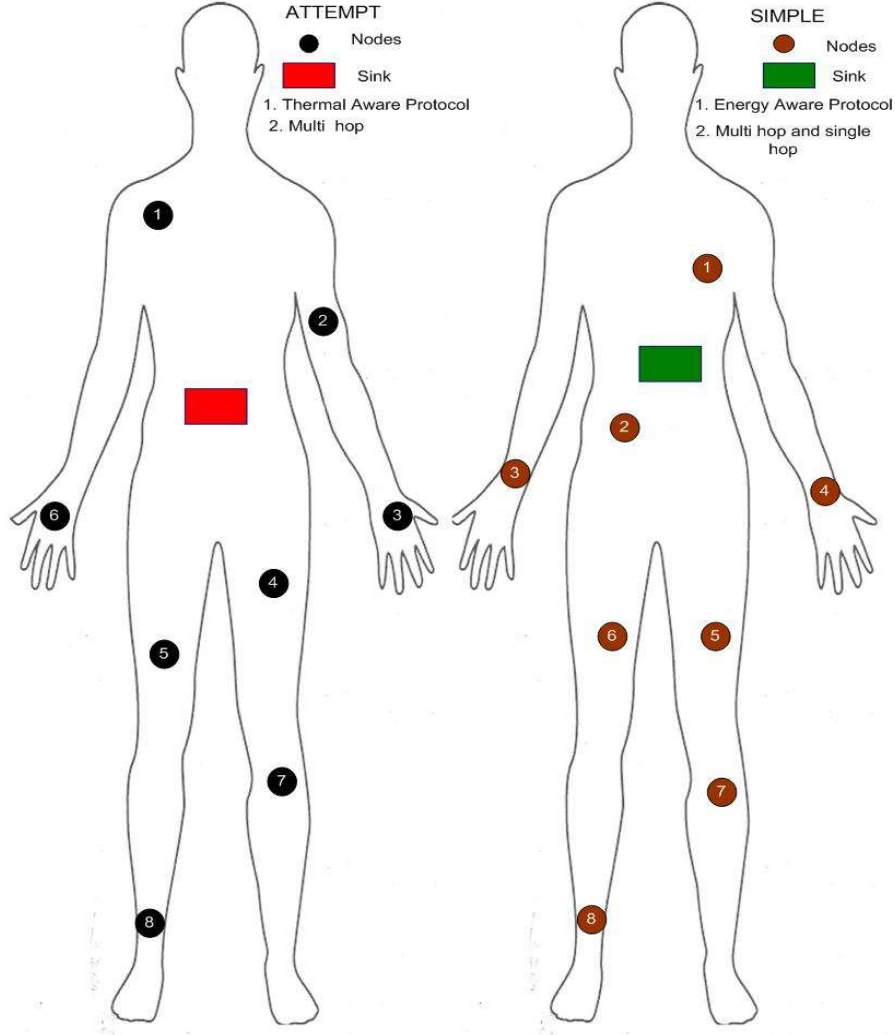}
\caption{Node deployment}
\end{center}
\end{figure*}

\subsection{Initial Phase}
In this phase, sink broadcast a short information packet which contains the location of the sink on the body. After receiving this control packet, each sensor node stores the location of sink. Each sensor node broadcasts an information packet which contains node ID, location of node on body and its energy status. In this way, all sensor nodes are updated with the location of neighbours and sink.
\subsection{Selection of next hop}
In order to save energy and to enhance network throughput, we proposed a multi hop scheme for WBAN. In this section, we present selection criteria for a node to become parent node or forwarder. To balance energy consumption among sensor nodes and to trim down energy consumption of network, SIMPLE protocol elects new forwarder in each round. Sink node knows the ID, distance and residual energy status of the nodes. Sink computes the cost function of all nodes and transmit this cost function to all nodes. On the basis of this cost function, each node decides whether to become forwarder node or not. If ${i}$ is number of nodes than cost function of ${i}$ nodes is computed as follows:\\

\begin{equation}
C.F({i})= \frac {d(i)} {R.E(i)}
\end{equation}

Where d$_{i}$ is the distance between the node ${i}$ and sink, R.E$_{i}$ is the residual energy of node ${i}$ and is calculated by subtracting the current energy of node from initial total energy. A node with minimum cost function is preferred as a forwarder. All the neighbor nodes stick together with forwarder node and transmit their data to forwarder. Forwarder node aggregates data and forward to sink. Forwarder node has maximum residual energy and minimum distance to sink; therefore, it consumes minimum energy to forward data to sink. Nodes for ECG and Glucose monitoring communicate direct to sink and do not participate in forwarding data.
\subsection{Scheduling}
In this phase, forwarder node assigns a Time Division Multiple Access (TDMA) based time slots to its children nodes. All the children nodes transmit their sensed data to forwarder node in its own scheduled time slot. When a node has no data to send, it switches to idle mode. Nodes wake up only at its transmission time. Scheduling of sensor nodes minimize the energy dissipation of individual sensor node.\\

\section{Performance Metrics}
We evaluated key performance metrics for proposed protocol. Definition of performance metrics is given in following subsections.

\subsubsection{Network lifetime}
 It represents the total network operation time till the last node die.\\
\subsubsection{Stability period}
 Stability period is the time span of network operation till the first node die. The time period after the death of first node is termed as unstable period.\\
\subsubsection{Throughput}
 Throughput is the total number of packets successfully received at sink.\\
\subsubsection{Residual Energy}
 In order to investigate the energy consumption of nodes per round, we consider residual energy parameter to analyze energy consumption of network.\\
\subsubsection{Path Loss}
 Path loss is the difference between the transmitted power of transmitting node and received power at receiving node. It is measured in decibels(dB).

\section{Simulation Results and Analysis}
To evaluate proposed protocol, we have conducted an extensive set of experiments using MATLAB R2009a. We studied the performance of the SIMPLE protocol by comparisons with the existing protocol M-ATTEMPT.
\subsection{Network life time}
Fig 2 shows the average network lifetime of proposed scheme. The proposed new cost function to elect forwarder node play an important role to balance the energy consumption among the sensor nodes. New forwarder in each round is selected based on computed cost function. Fig 2 clearly depicts that the proposed protocol has longer stability period. This is expected, due to the appropriate selection of new forwarder in each round. Hence, each node consumes almost equal energy in each round and all the nodes die almost at the same time. In M-ATTEMPT, as temperature of forwarder nodes increases, nodes select alternate longer path which consumes more energy. Hence, these nodes die early. Our proposed protocol achieves 31\% more stability period and 0.4\% longer network lifetime.

\begin{figure}[!h]
\begin{center}
\includegraphics[scale=0.5]{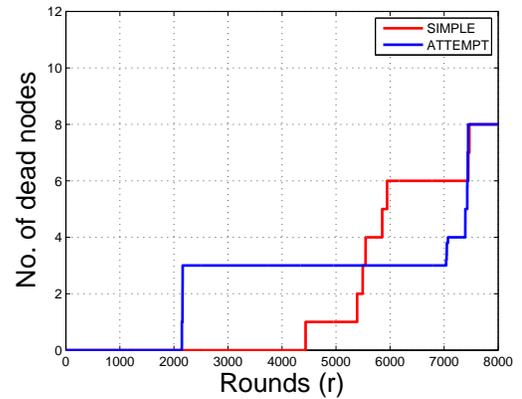}
\caption{Analysis of network lifetime}
\end{center}
\end{figure}

\subsection{Throughput}
Throughput is the successful packet received at the sink. As WBAN has critical and important data of patient, so it requires a protocol which has minimum packet drop and maximum successful data received at sink. SIMPLE protocol achieves high throughput than M-ATTEMPT, as shown in fig 3. Number of packets send to sink depends on the number of alive nodes. More alive nodes send more packets to sink which increases the throughput of network. The stability period of M-ATTEMPT is shorter than SIMPLE protocol which means number of packets sent to sink decreased. Hence, throughput of M-ATTEMPT decreased. On the hand, SIMPLE protocol achieves high throughput due to longer stability period.\\

\begin{figure}[!h]
\begin{center}
\includegraphics[scale=0.5]{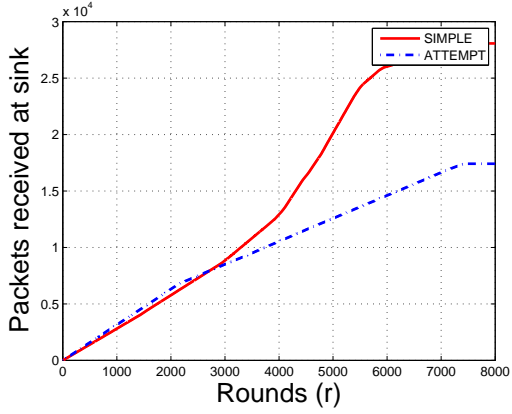}
\caption{Analysis of Throughput}
\end{center}
\end{figure}

\subsection{Residual energy}
The average energy of network consumed in each round is presented in fig 4. The proposed model use multi hop topology, in which each farthest node transmits its data to sink through a forwarder node. Forwarder node is elected using aforementioned cost function. Selection of appropriate forwarder in each round contributes to save energy. To transfer packets to sink, our multi hop topology use different forwarder node in each round, this restricts over loading of particular node. Simulation results show that SIMPLE protocol consumes minimum energy till 70\% of simulation time. It means, in stability period, more nodes have enough energy and they transmit more data packet to sink. It also improves the throughput of the network. On the other hand, in M-ATTEMPT, some nodes exhaust early due to heavy traffic load.\\

\begin{figure}[!h]
\begin{center}
\includegraphics[scale=0.5]{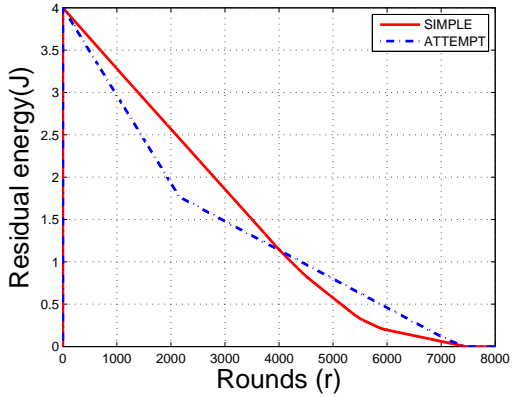}
\caption{Analysis of remaining energy}
\end{center}
\end{figure}

\subsection{Path Loss}

Fig 5 presents the path loss of different sensors. Path loss is a function of frequency and distance. Path loss shown in figure 5 is function of distance. It is calculated from its distance to sink with constant frequency 2.4GHz. We use path loss coefficient 3.38 and 4.1 for standard deviation $\sigma$. Proposed multi hop topology reduces the path loss as shown in figure 5. It is due to the fact that multi hop transmission reduces the distance, which leads to minimum path loss. Fig 5 represents the results of both topologies. Initially SIMPLE protocol performs well. However, after 2000 rounds, path loss of M-ATTEMPT dramatically decreased because some nodes of M-ATTEMPT topology die. Minimum number of alive nodes has minimum cumulative path loss. As our proposed protocol has longer stability period and more alive nodes has more cumulative path loss.\\

\begin{figure}[!h]
\begin{center}
\includegraphics[scale=0.5]{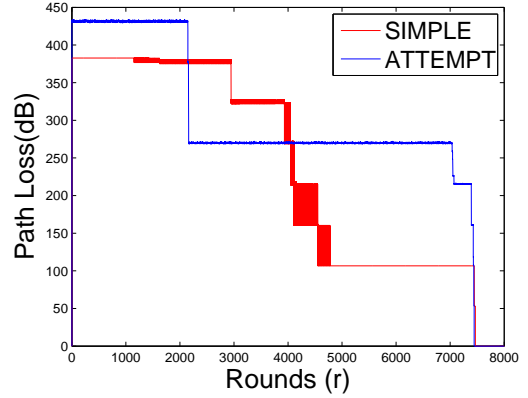}
\caption{ network path loss }
\end{center}
\end{figure}

\subsection{Path loss model}
Path loss represents the signal attenuation and is measured in decibels (dB). Signal power is also degraded by Additive White Gaussian Noise
(AWGN)~\cite{manzoor2012noise}. Path loss is the difference between the transmitted power and received power whereas antenna gain may or may not be considered. Path loss occurs due to the increasing surface area of propagating wave front. Transmitting antenna radiates power outward and any object between transmitter and receiver causes destruction of radiated signal. In WBAN, different human postures, movement of body, hands and cloths, affects the transmitted signal. Path loss is related to the distance and frequency and expressed as~\cite{javaid2013ubiquitous}.

\begin{equation}
  PL(f,d)= PL(f) \times PL(d)
\end{equation}

The relation of frequency with path loss is expressed as

\begin{equation}
 \sqrt{PL(f)} \propto f^k
\end{equation}

Where k is frequency dependent factor and it is related to the geometry of the body.
The relation of distance with path loss is given as

\begin{equation}
  PL(f,d)= PL_{o} + 10 \emph{n} log _{10} \frac {d}{do}+ X \sigma
\end{equation}

Where $PL$ is received power, $d$ is the distance between transmitter and receiver, d$_{o}$ is the reference distance, n is the path loss coefficient and its value depends on the propagation environment. In free space its value is 2, for WBAN, $n$ varies from 3-4 for line of sight (LOS) communication and 5-7.4 for non line of sight (NLOS) communication. $X$ is gaussian random variable and ${\sigma}$ is standard deviation~\cite{rappaport2002principles}. PL$_{o}$ is received power at reference distance d$_{o}$ and it is expressed as:

\begin{equation}
  PL_{o} =  10 log _{10} \frac {(4\pi \times d \times f)}{c}^2
\end{equation}

Where ${f}$ is frequency, ${c}$ speed of light and d is distance between transmitter and receiver. The value of reference distance d$_{o}$ is 10cm. In reality it is difficult to predict strength of signal between transmitter and receiver boundary. To solve this issue, we use a deviation variable X${\sigma}$.

\section{Conclusion}
In this paper, we propose a mechanism to route data in WBANs. The proposed scheme use a cost function to select appropriate route to sink. Cost function is calculated based on the residual energy of nodes and their distance from sink. Nodes with less value of cost function are elected as parent node. Other nodes become the children of that parent node and forward their data to parent node. Two nodes for ECG and Glucose monitoring forward their data direct to sink as they are placed near sink, also these two nodes can not be elected as parent node because both sensor node has critical and important medical data. It is not required that these two node deplete their energy in forwarding data of other nodes. Our simulation results shows that proposed routing scheme enhance the network stability time and packet delivered to sink. Path loss is also investigated in this protocol and in future work, we will implement Expected Transmission Count (ETX) link metrics as demonstrated in~\cite{javaid2009performance}~\cite{dridi2010performance}~\cite{dridi2009ieee}.


\end{document}